\documentclass[]{emulateapj}
\usepackage{amsmath,amstext}
\usepackage[breaklinks,colorlinks,citecolor=blue,linkcolor=magenta]{hyperref} 
\usepackage[all]{hypcap} 
\usepackage{float}
\usepackage{multirow}
\usepackage{lineno}

\def \nustar {{\em NuSTAR }}
\def \xmm {{\em XMM-Newton }}
\def \suzaku {{\em Suzaku }}
\def \sax {{\em Beppo-SAX }}

\def \iras {{IRAS\,05189--2524 }}

\newcommand{\Msun}      {\mbox{$M_{\mathord\odot}$}}

\newcommand{\Lsun}      {\mbox{$L_{\mathord\odot}$}}

\shorttitle{\iras}
\shortauthors{Xu et al.}

\begin{document}

\title{Evidence for relativistic disk reflection in the Seyfert 1h galaxy/ULIRG \iras observed by NuSTAR and XMM-Newton}

\author{Yanjun Xu\altaffilmark{1}, Mislav Balokovi\'{c}\altaffilmark{1}, Dominic J. Walton\altaffilmark{1,2,3}, Fiona A. Harrison\altaffilmark{1}, Javier A. Garc\'ia\altaffilmark{1,4,5,8}, Michael J. Koss\altaffilmark{6,7,9}}

\altaffiltext{1}{Cahill Center for Astronomy and Astrophysics, California Institute of Technology, Pasadena, CA 91125, USA}
\altaffiltext{2}{Jet Propulsion Laboratory, California Institute of Technology, Pasadena, CA 91109, USA}
\altaffiltext{3}{Institute of Astronomy, University of Cambridge, Madingley Road, Cambridge CB3 0HA, UK}
\altaffiltext{4}{Harvard-Smithsonian Center for Astrophysics, 60 Garden St., Cambridge, MA 02138 USA}
\altaffiltext{5}{Remeis Observatory \& ECAP, Universit\"at Erlangen-N\"urnberg, Sternwartstr.~7, 96049 Bamberg, Germany}
\altaffiltext{6}{Institute for Astronomy, Department of Physics, ETH Zurich, Wolfgang-Pauli-Strasse 27, CH-8093 Zurich, Switzerland}
\altaffiltext{7}{Institute for Astronomy, University of Hawaii, 2680 Woodlawn Drive, Honolulu, HI 96822, USA}
\altaffiltext{8}{Alexander von Humboldt Fellow.}
\altaffiltext{9}{SNSF Ambizione Fellow.}

\begin{abstract}
We present a spectral analysis of the \nustar and \xmm observations of the Seyfert 1h galaxy/ULIRG {IRAS\,05189--2524} taken in 2013. We find evidence for relativistic disk reflection in the broadband X-ray spectrum: a highly asymmetric broad Fe K$\alpha$ emission line extending down to 3\,keV and a Compton scattering component above 10\,keV. Physical modeling with a self-consistent disk reflection model suggests the accretion disk is viewed at an intermediate angle with a super-solar iron abundance, and a mild constraint can be put on the high-energy cutoff of the power-law continuum. We test the disk reflection modeling under different absorption scenarios. A rapid black hole (BH) spin is favored, however we cannot place a model-independent tight constraint on the value. The high reflection fraction ($R_{\rm ref} \simeq$ 2.0--3.2) suggests the coronal illuminating source is compact and close to the BH (lying within 8.7 $R_{\rm g}$ above the central BH), where light-bending effects are important. 
\end{abstract}

\maketitle

\section{Introduction}
The hard X-ray continuum emission of active galactic nuclei (AGNs) is believed to be produced in the corona via thermal Comptonization of ultraviolet (UV) radiation from the accretion disk \citep[e.g.,][]{haa93}. This continuum emission is subsequently reprocessed by dense material either from the accretion disk or distant structures such as the dusty torus or the broad line region (BLR), producing the characteristic reflected X-ray spectrum \citep[e.g.,][]{geo91}. Fluorescent spectral lines, line edges and a Compton hump are typical features of the reflected X-ray spectrum. Fe K$\alpha$ is the most prominent spectral line in the X-ray band  due to its high cosmic abundance and fluorescent yield. The line emission from the innermost regions of the accretion disk is relativistically broadened by the combined effects of Doppler shifts and gravitational redshift \citep{fab89, lao91}, which skew the line profile into an extended red wing. Assuming the accretion disk extends down to the innermost stable orbit (ISCO), which is determined by the spin of the black hole (BH), a relativistically broadened Fe K$\alpha$ line could be used as a probe of the BH spin.

So far, broad Fe K$\alpha$ lines have been detected in many AGNs \citep[e.g.][]{tan95, nan07, ris13, wal13}; these are mostly Seyfert\,1--1.5 galaxies with only a few absorbed Seyfert 2 galaxies showing such features: e.g., IRAS 18325--5926 \citep{iwa96}, IRAS 00521--7054 \citep{tan12}. \xmm and \suzaku observations of IRAS\,00521--7054 revealed a possibly strong and broad Fe K$\alpha$ line favoring a high BH spin, however the broad line could also be artificially produced by a combination of partial covering absorbers \citep{tan12, ric14}. With high-quality broadband X-ray data and multi-epoch observations, BH spins have been robustly constrained in some nearby obscured Seyferts \citep[e.g., Seyfert 1.9 galaxy NGC 1365,][]{ris13, wal14}, but the measurement is still difficult for these obscured objects.

\iras is an ultraluminous infrared galaxy (ULIRG) with log\,($L_{\rm IR}/\Lsun)=12.16$ \citep{u12} and hosts a Compton-thin AGN at redshift $z=0.0426$. While originally classified as Seyfert~2 \citep{vei95}, broad polarized Balmer lines were detected \citep{you96} and near-infrared spectroscopy revealed broad Paschen lines \citep{vei99b, sev01}, indicating the existence of a hidden BLR. Therefore, it has been more accurately classified as a hidden broad line Seyfert 1 galaxy (Seyfert 1h galaxy/S1h) \citep{ver06}, which defines the kind of Seyfert 2s having the spectra of Seyfert 1s in the polarized light. 

In the X-ray band, \iras is one of the most luminous ULIRGs in the sky, which are typically merger systems undergoing rapid star-forming activities. \iras is likely a late stage merger of two spiral galaxies \citep{san88}. Its X-ray emission is dominated by the central AGN rather than starburst activities \citep{vei99b}. Iron emission has been detected in \iras from previous X-ray observations by \sax \citep{sev01} and \suzaku \citep{ten09}, but the general line profile was not resolved. The source went through a major spectral change during the 2006 \suzaku observation when the 2--10 keV flux dropped by a factor of $\sim$30, which could be explained by an increase in the absorption column density or the change of the intrinsic AGN luminosity \citep{ten09}. At a galactic scale, high-velocity outflows of ionized \citep{wes12, bel13}, neutral \citep{rup05, ten13, rup15} and molecular \citep{vei13} gas have been observed.

Outflows have been recognized as a signature of AGN feedback, which is believed to play an important role in the co-evolution of AGNs and their host galaxies. Evidence for multi-phase outflowing gas has been observed in AGNs on a wide rage of spatial scales from that of the central accretion disk to large-scale molecular outflows. If outflows at different scales are all AGN-driven, linking them could help unveil the expansion mechanism of the wind, such studies have been conducted on ULIRGs IRAS\,11119+3257 \citep{tom15} and Mrk\,231 \citep{fer15}. In the X-ray band, there is growing evidence for ultra-fast outflows (UFOs) with the typical velocities of $\sim$0.1--$0.2\,c$, detected by fitting blue-shifted\,Fe K-shell absorption lines between 7 and 10\,keV \citep[e.g.,][]{pou03, tom10, pou13}. In the case of PDS 456, a P-Cygni-like profile was observed, demonstrating a powerful wide angle outflow \citep{nar15}.

In this paper, we report evidence for relativistic disk reflection in \iras based on analysis of the broadband X-ray spectrum from \nustar and {\em XMM-Newton}. In Section~\ref{sec:data}, we describe the observations and the data reduction. Section~\ref{sec:spec} provides the details of our spectral fitting. We present a discussion of the results in Section~\ref{sec:dis} and summarize our results in Section~\ref{sec:con}.

\section{DATA REDUCTION}
\label{sec:data}
\iras was observed with \nustar \citep{har13} and \xmm \citep{jan01} in 2013 as a part of the \nustar ULIRG survey \citep{ten15} and the \nustar survey of Swift/BAT AGN (Balokovi\'{c} et al., in prep.). \nustar observed the source at two epochs: on 2013 February~20 (OBSID 60002027002) and on 2013 October~2 (consecutive OBSIDs 60002027004 and 60002027005) with the exposures of 23.1\,ks, 25.4\,ks and 8.3\,ks, respectively. The \xmm observation (OBSID 0722610101) with an exposure time of 37.8\,ks was coordinated with the second \nustar observation. The dataset was previously analysed in the \nustar ULIRG survey \citep{ten15}, with the focus on uncovering the obscuration levels and intrinsic luminosities of several nearby ULIRGs. We reduced the data following the data processing description in \cite{ten15}. 

\subsection{NuSTAR} 
The \nustar data were processed using v.1.6.0 of the NuSTARDAS pipeline with \nustar CALDB v20160731. For each observation, the source spectra were extracted at the position of \iras within the radius of 60$\arcsec$. Corresponding background spectra were extracted from source-free areas on the same chip using polygonal regions. As discussed in \cite{ten15}, the \nustar data show minor count rate variability ($\sim$20\%) and no obvious change in spectral shape between the two epochs. Therefore, we combined the three \nustar observations using the {\small ADDSPEC} script in HEASoft~v6.19 to maximize the signal-to-noise ratio (S/N) of the spectra. We binned the data taking into account the background level to provide a nearly constant S/N over the NuSTAR bandpass, with a minimum S/N of 3, and median S/Ns of 5.5 and 5.3 for the two focal plane modules (FPMA and FPMB).

\subsection{XMM-Newton}
The \xmm observation was reduced with the \xmm Science Analysis System v14.0.0 following standard procedures. The raw event files were filtered using {\small EPCHAIN} and {\small EMCHAIN} to produce cleaned event lists for each of the EPIC-pn \citep{str01} and EPIC-MOS \citep{tur01} detectors, respectively. Science products were then produced using {\small XMMSELECT}, considering only single and double events for EPIC-pn, and single to quadruple events for EPIC-MOS. We extracted the source from a circular region of radius 40$''$, while background was estimated from a large region of blank sky on the same detector as IRAS\,05189--2524. Instrumental response files were generated with {\small RMFGEN} and {\small ARFGEN} for each detector. The good exposure is $\sim$31\,ks for EPIC-pn and $\sim$36\,ks for each of the two EPIC-MOS detectors. After performing the reduction separately for MOS1 and MOS2, we combined these data into a single spectrum using {\small ADDASCASPEC}. The MOS and PN spectra were grouped to have 25 counts per bin.

\section{SPECTRAL ANALYSIS}
\label{sec:spec}
We model the time-averaged \nustar and \xmm spectra jointly in XSPEC v12.9.0 \citep{arn96} using the $\chi^{2}$ statistics. All uncertainties of the spectral parameters are reported at the 90\% confidence level. Cross-normalization constants are allowed to vary freely for \nustar FPMB and \xmm MOS and PN, and assumed to be unity for FPMA. The values are within $\sim$10\% of unity, as expected from \cite{mad15}. For all the spectral fitting, we take into account the Galactic absorption by modifying the spectrum with a {\tt TBabs} absorption model \citep{wil00}. The Galactic column density is set to  $N_{\rm H,gal}=1.66\times10^{20}$ cm$^{-2}$ \citep{kal05}. 

\capstartfalse
\begin{deluxetable}{cll}[]
\tablewidth{\columnwidth}
\tablecolumns{3}
\tabletypesize{\scriptsize}
\tablecaption{XSPEC Models \label{tab:tab1}}
\tablehead{
\colhead{Model} & \colhead{XSPEC Components}  & \colhead{Spectral Range}} 
\startdata
 1      &  zpcfabs*powerlaw + mekal   & 0.3--30\,keV\\
\noalign{\smallskip}
 2      & zpcfabs*(pexrav + zgauss) + mekal & 0.3--30\,keV\\
\noalign{\smallskip}
 3a    &  zpcfabs*relxilllp & 2--30\,keV\\
\noalign{\smallskip}
 3b    & (partcov*tbnew\_feo)*relxilllp & 2--30\,keV\\
\noalign{\smallskip}
 4a    & zpcfabs*{\scriptsize XSTAR}*relxilllp & 2--30\,keV\\
\noalign{\smallskip}
 4b  &  (partcov*tbnew\_feo)*{\scriptsize XSTAR}*relxilllp & 2--30\,keV
\enddata
\end{deluxetable}

We begin our spectral analysis of the broadband X-ray spectrum considering the band from 0.3\,keV to 30\,keV, as it becomes background-dominated above 30\,keV. We first fit the spectrum with a power-law model subject to one partial covering neutral absorber modeled with {\tt zpcfabs}, plus a {\tt mekal} line emission model with its elemental abundances fixed at solar values (model~1, XSPEC models are listed in Table~\ref{tab:tab1}). The {\tt mekal} component is used to model the thermal plasma contribution from the host galaxy. We find the best fit (reduced chi-square $\chi^{2}/{\rm d.o.f.}=1425.8/1049$) with a photon index $\Gamma=2.29\pm0.04$ and a {\tt mekal} temperature $kT=0.092^{+0.010}_{-0.005}$\,keV. The covering fraction of the neutral absorber is well constrained as 98.89\%, therefore contribution from the scattered power-law component is negligible above $\sim$2\,keV. For simplicity, we fix the value at 99\% from here on.

\begin{figure}
\centering
\includegraphics[width=0.49\textwidth]{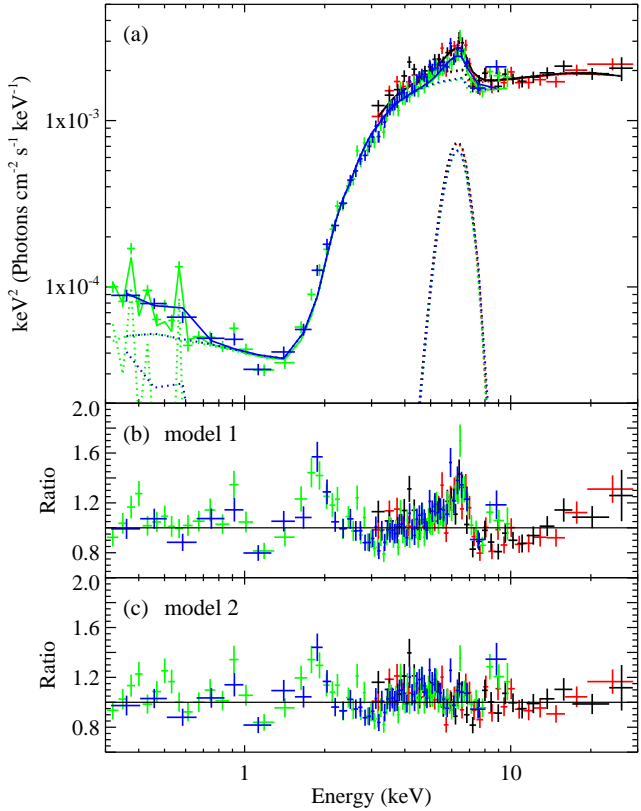}
\caption{(a) \xmm PN (green), MOS (blue), \nustar FPMA (red) and FPMB (black) spectra of {IRAS\,05189--2524}, with the best-fit phenomenological reflection model (model 2).  (b) Data/model ratio after fitting an absorbed power-law plus a soft {\tt mekal} component to the 0.3--30\,keV spectrum (model 1). (c) Residuals of the phenomenal reflection model (model 2).  The data are rebinned for display clarity. 
\label{fig:fig1}}
\end{figure}

\begin{figure}
\centering
\includegraphics[width=0.49\textwidth]{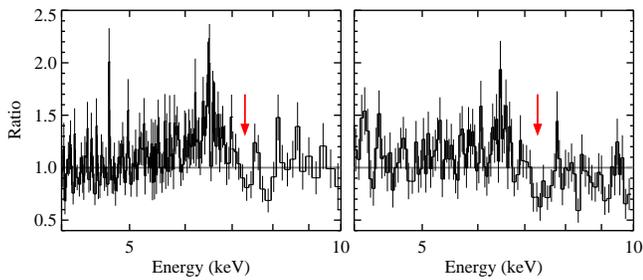}
\caption{Possible absorption features between 7\,keV and 8\,keV marked by arrows in the residual plots of the absorbed power-law model (model 1). Left panel: \xmm PN spectrum; right panel: \nustar FPMB spectrum.  
\label{fig:fig2}}
\end{figure}

The data/model ratio plot shows strong signatures of disk reflection (Figure~\ref{fig:fig1}, panel b): a broad iron line peaking around 6.7\,keV in the source rest-frame with a skewed shape extending down to $\sim$3\,keV and residuals above 10\,keV from the Compton reflection. In addition, there are possible absorption features of blue-shifted Fe K lines in both the \nustar and the \xmm spectra, with the most evident absorption troughs lying in the vicinity of the Fe\,K-edge around 7.3\,keV (Figure~\ref{fig:fig2}). 

\capstartfalse
\begin{deluxetable}{clll}
\tablewidth{\columnwidth}
\tablecolumns{4}
\tabletypesize{\scriptsize}
\tablecaption{Spectral Fitting of \iras: Part I \label{tab:tab2}}
\tablehead{
\colhead{Component} & \colhead{Parameter} & \colhead{Model 1} & \colhead{Model 2} } 
\startdata
{\textsc{zpcfabs}}  & $N_{\rm H}$ ($\rm \times10^{22}~cm^{-2}$) &$8.82^{+0.27}_{-0.26}$ &$8.30\pm0.21$ \\
\noalign{\smallskip}
                    & $f_{\rm abs} (\%)$ &$98.89^{+0.08}_{-0.09}$ & $99^{\star}$    \\
\noalign{\smallskip}
\hline	
\noalign{\smallskip}		
{\textsc{power-law}}  & $\Gamma$  &$2.29\pm0.04$  & \nodata    \\
\noalign{\smallskip}
{\textsc{mekal}}    & {kT~(keV)}  &$0.092^{+0.010}_{-0.005}$ &$0.087^{+0.006}_{-0.009}$ \\
\noalign{\smallskip}
\hline
\noalign{\smallskip}
{\textsc{pexrav}}      &   $\Gamma$  & \nodata  & $2.47\pm0.04$    \\
\noalign{\smallskip}
                 &   $R_{\rm ref}$  & \nodata  & $1.48^{+0.53}_{-0.44}$ \\
\noalign{\smallskip}
\hline
\noalign{\smallskip}
{\textsc{zgauss}}      &   {LineE~(keV)}  & \nodata    & $6.40^{\star}$       \\
\noalign{\smallskip}
            &   $\sigma$~(keV)  & \nodata  & $0.71^{+0.13}_{-0.11}$    \\
\noalign{\smallskip}
\hline
\noalign{\smallskip}
            &   $\chi^2/{\rm d.o.f}$  & 1425.8/1049 & 1194.8/1047 
\enddata
\tablecomments{
Model 1 and model 2 fit the 0.3--30\,keV broadband spectrum from \nustar and \xmm. Parameters with $\star$ are fixed values.
}
\end{deluxetable}

\subsection{Phenomenological Reflection Model}
We then test fitting the reflection features with a phenomenological model by adding a neutral reflection component {\tt pexrav} \citep{mag95} and one Gaussian emission line with variable width using the model {\tt zgauss} (model~2, see table~\ref{tab:tab1}). We freeze the centroid of the Gaussian emission line at 6.4\,keV, the disk inclination angle at the default value ({\tt cos\,i} = 0.45), the iron abundance at solar and include no high-energy cutoff ({\tt foldE} = 0), as this simple model is not sensitive to these parameters. The fit is improved considerably with $\Delta \chi^{2}/\Delta{\rm d.o.f} = -231/-2$. A broad Gaussian line with line width $\sigma = 0.71^{+0.13}_{-0.11}$\,keV and equivalent width (EW) $\simeq 0.60$\,keV is required to provide a decent fit for the iron line emission feature in the spectrum. However, a clear excess still remains in the residuals between 3 and 6\,keV as would be expected from the red wing of a relativistically broadened iron line (Figure~\ref{fig:fig1}, panel c). Model 2 also gives a very steep power-law of $\Gamma=2.47\pm0.04$. We do not attempt to interpret the intrinsic continuum steepness at this point, as this is only a phenomenological model.

We note that for both the simple absorbed power-law and the phenomenological reflection model, at least two extra emission features are present in the residuals around 0.9 and 1.9\,keV. The 0.9\,keV excess might be associated with O Ly$\alpha$ or Fe--L shell emissions, and the feature around 1.9\,keV could be caused by ionized Si emission lines. The excesses cannot be adequately fitted with simple Gaussian emission line models, as they could be a complicated combination of ionized absorption, over-abundant metal lines and X-ray contribution from the high mass X-ray binary population in the star-forming regions. X-ray point-source emission correlates with galaxy star forming rate (SFR) in star-forming galaxies \citep{leh10}. At an SFR of $\sim80\,\Msun\,\rm\,yr^{-1}$ \citep{wes12}, the 2--10 keV luminosity from the binary population is estimated to be $1.4\times10^{41}\rm\,ergs\,s^{-1}$, about 1\% of the total luminosity, which is enough to dominate the soft emission of {IRAS\,05189--2524}. Therefore, the soft part of the spectrum likely helps little to constrain the X-ray continuum from the AGN. In order to avoid the possible situation where the soft emission from the host galaxy is driving our modeling of the disk reflection features, we ignore the data below 2 keV from here on.
                
\subsection{Self-consistent Disk Reflection Models}
In \cite{ten15}, the authors attempted to explain the reflection features observed in \iras by neutral reflection from the torus, which is commonly the case for Compton-thick AGNs. However, they found neither the {\tt MYTorus} \citep{mur09} nor {\tt BNTorus} model \citep{bri11} could provide an adequate fit for the broadband spectrum, which ruled out the possibility of distant reflection. In order to explore the relativistic reflection hypothesis, we replace the {\tt pexrav} and {\tt zgauss} components in the phenomenological model with the self-consistent relativistic reflection model {\tt relxill} \citep{dau14, gar14} (model~3a, see Table~\ref{tab:tab1}). We use {\tt relxilllp} in the {\tt relxill} model family, instead of using an empirical emissivity law, the emissivity is calculated directly in the lamp post geometry \citep{dau13}. {\tt Relxilllp} calculates a self-consistent reflection fraction. This could help us constrain the BH spin by shrinking the parameter space, as results with unphysical solutions of low BH spins with high reflection fractions can be ruled out by the assumption of a lamp-post geometry (for discussion, see \citealt{dau14}). In order to measure the BH spin, we assume the inner disk radius extends down to the ISCO. We fix the outer radius at the default value of 400\,$R_{\rm g}$ ($R_{\rm g} \equiv$ GM/c$^{2}$ is the gravitational radius) as the model is not sensitive to this parameter, and the cutoff energy is allowed to vary in the fitting.

\begin{figure}
\centering
\includegraphics[width=0.49\textwidth]{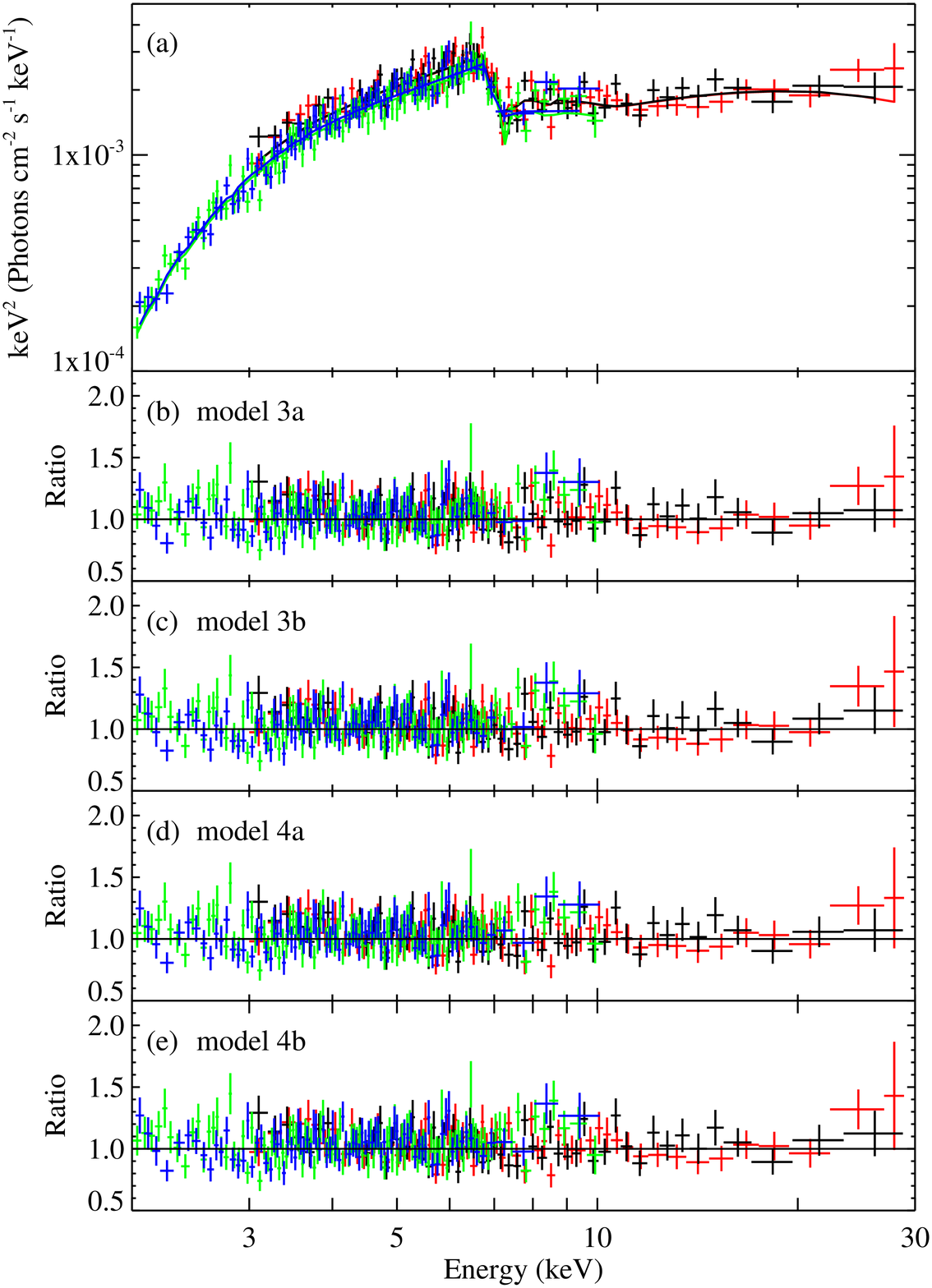}
\caption{(a) \xmm and \nustar spectra of {IRAS\,05189--2524} with the relativistic disk reflection model 4b.  (b)--(e)  Data/model ratio plots after fitting for the disk reflection component under four different absorption scenarios: (b) one neutral absorber with a solar iron abundance; (c) one neutral absorber with a variable iron abundance; (b) one neutral absorber with a solar iron abundance + one ionized absorber; (c) one neutral absorber with a variable iron abundance + one ionized absorber. The data are rebinned for display clarity.  See Table~\ref{tab:tab1} for the list of XSPEC models.
\label{fig:fig3}} 
\end{figure}

The {\tt relxilllp} model yields a good fit to the 2--30\,keV spectrum with $\chi^{2}/$d.o.f.$=862.0/859$ and no obvious excess in the residuals (Figure~\ref{fig:fig3}, panel b). The intrinsic X-ray continuum is steep, with the power-law photon index $\Gamma=2.04^{+0.02}_{-0.03}$. It is in the upper range of the photon index distribution observed in local Seyfert galaxies \citep[e.g.][]{nan94,win09,riv13}, but less extreme than the previously published values for IRAS\,05189--2524: e.g, $\Gamma = 2.68^{+0.30}_{-0.13}$ \citep{ten09}, $\Gamma =2.51 \pm 0.02 $ \citep{ten15}. Albeit lacking high quality data above 30\,keV, the model is able to put a modest constraint on the high-energy cutoff $E_{\rm cut} = 55^{+10}_{-7}$\,keV, as discussed in \cite{gar15a}, such sensitivity could come partly from the low energy part of the reflection spectrum ($<$ 3\,keV). The location of the primary radiation source is found to be close to the disk $h=2.39^{+0.80}_{-0.32}$\,$R_{\rm g}$, with a rapid BH spin (dimensionless spin parameter $a=0.96^{+0.02}_{-0.03}$), and the self-consistent reflection fraction derived by the model is high ($R_{\rm ref}$ = 3.24). The best-fit results also find an intermediate inclination of $\theta = 42\pm2$~degrees and an ionization parameter\footnote{The ionization parameter $\xi=4\pi F_{\rm x}/n$, where $F_{\rm x}$ is the ionizing flux and $n$ is the gas density.} log$~\xi$$=3.02\pm0.02~\rm\,erg~cm~s^{-1}$ for the accretion disk. We note that the iron abundance $A_{\rm Fe}$ is unusually high with best-fit value hitting the upper bound of the model of 10. But if we fix the iron abundance at solar in the {\tt relxilllp} model, the fitting degrades considerably ($\Delta \chi^{2} \sim 130$) with a visible excess in the Fe K band.

\capstartfalse
\begin{deluxetable*}{clllll}
\tablewidth{\textwidth}
\tablecolumns{6}
\tabletypesize{\scriptsize}
\tablecaption{Spectral Fitting of \iras: Part II Self-consistent Disk Reflection Models \label{tab:tab3}}
\tablehead{
\colhead{Component} & \colhead{Parameter} &\colhead{Model 3a} &  \colhead{Model 3b} & \colhead{Model 4a} & \colhead{Model 4b}  } 
\startdata
{\textsc{zpcfabs}}  & $N_{\rm H}$ ($\rm \times10^{22}~cm^{-2}$) &$7.17^{+0.17}_{-0.16}$ & $4.20^{+0.21}_{-0.27}$ & $7.15^{+0.62}_{-0.53}$ & $4.33^{+0.98}_{-0.33}$ \\
\noalign{\smallskip}
                    & $f_{\rm abs} (\%)$  &$99^{\star}$ & $99^{\star}$ & $99^{\star}$  & $99^{\star}$      \\
\noalign{\smallskip}
                 &  {$A_{\rm Fe}$} & \nodata & $5.5^{+2.5}_{-1.1}$   & \nodata   & $5.0^{+u}_{-0.7}$  \\
\noalign{\smallskip}
\hline	
\noalign{\smallskip}
{\textsc{xstar}}       & $N_{\rm H}$ ($\rm \times10^{22}~cm^{-2}$)   & \nodata      & \nodata    & $0.27^{+0.37}_{-0.27}$   & $1.44^{+1.56}_{-1.44}$   \\
\noalign{\smallskip}
            &   log~$\xi$ ($\rm erg~cm~s^{-1}$)  & \nodata  & \nodata     & $3.20^{+1.26}_{-0.68}$    & $3.17^{+0.28}_{-0.39}$   \\
\noalign{\smallskip}
		  &   $V_{\rm out}$ (c)   & \nodata     & \nodata      & $0.13^{+0.02}_{-0.06}$     & $0.12^{+0.02}_{-0.01}$ \\
\noalign{\smallskip}
            &   {$A_{\rm Fe}$}  & \nodata    & \nodata     & $10.0^{+u}_{-3.3}$     & $5.0^{+u}_{-0.7}$    \\                         
\noalign{\smallskip}
\hline
\noalign{\smallskip}
{\textsc{relxilllp}}  &h (GM/c$^{2}$)  & $2.39^{+0.80}_{-0.32}$  & $4.70^{+4.04}_{-1.99}$  & $3.00^{+1.48}_{-0.87}$ & $4.08^{+2.16}_{-1.60}$    \\
\noalign{\smallskip}                      
				&   a (cJ/GM$^{2}$)   & $0.96^{+0.02}_{-0.03}$  & $0.96^{+u}_{-0.94}$ & $0.96^{+u}_{-0.07}$ & $0.98^{+u}_{-0.61}$    \\
\noalign{\smallskip}                         
				&   $\theta$ ($^\circ$)   & $42\pm2$  & $44^{+2}_{-5}$  & $41^{+5}_{-2}$  & $51^{+4}_{-5}$ \\
\noalign{\smallskip}                         
				&   $\Gamma$   & $2.04^{+0.02}_{-0.03}$ & $1.79^{+0.17}_{-0.14}$   & $2.00^{+0.17}_{-0.19}$  & $1.74^{+0.18}_{-0.17}$ \\
\noalign{\smallskip}                        
                    &   log~${\xi}$ ($\rm erg~cm~s^{-1}$) & $3.02\pm0.02$   &  $3.12^{+0.22}_{-0.11}$  & $3.06\pm0.07$  & $3.23^{+0.14}_{-0.18}$   \\
\noalign{\smallskip}                          
				&   {$A_{\rm Fe}$}  &$10.0^{+u}_{-1.3}$   & $5.5^{+2.5}_{-1.1}$  & $10.0^{+u}_{-3.3}$   & $5.0^{+u}_{-0.7}$     \\
\noalign{\smallskip}                          
				&   {$E_{\rm cut}$~(keV)}  & $55^{+10}_{-7}$   & $34^{+21}_{-10}$  & $59^{+111}_{-24}$    & $33^{+21}_{-10}$    \\
\noalign{\smallskip}  
\hline	
\noalign{\smallskip}		
            &   $\chi^2/{\rm d.o.f}$  & 862.0/859 & 860.6/859  & 854.8/856  & 854.7/856    
\enddata
\tablecomments{
Model 3--4 fit the broadband spectrum with the part below 2.0\,keV ignored.
Parameters with $\star$ are fixed values. 
Iron abundances ($A_{\rm Fe}$) are linked during the spectral fitting.
Values with uncertainty marked with $+u$ denote 90\% confidence limits in excess of the upper bound of the model.
}
\end{deluxetable*}

A super-solar metal abundance is not uncommon among ULIRGs \citep[e.g.,][]{rup08}. In the case of {IRAS\,05189--2524}, based on the diagnostics of [N II]/H$\alpha$, [S II]/H$\alpha$ and [O I]/H$\alpha$ line ratios, an uncommonly high metallicity ($>4 {Z_{\odot}}$)\footnote{Only an upper limit was put on the metallicity because the line ratios were outside the range predicted by the photoionization, shock or AGN models used in the paper.} is estimated for the outer galactic disk, or else some peculiar ionization effects are needed to explain the line ratio measurements \citep{wes12}. If the metallicity is indeed high on the galactic scale in {IRAS\,05189--2524}, it is more physically consistent to use neutral absorption models with super-solar abundances in our spectral fitting, as neutral absorbers normally reside far from the central BH.

In an attempt to further constrain the iron abundance, we tie the iron abundances of the neutral absorber and the disk reflection component by replacing the {\tt zpcfabs} model with an improved model for neutral absorption {\tt tbnew\_feo}\footnote{http://pulsar.sternwarte.uni-erlangen.de/wilms/research/tbabs/} convolved with the partial covering model {\tt partcov} (model~3b, see Table~\ref{tab:tab1}). Both Fe and O abundances are variable parameters in the {\tt tbnew\_feo} model. We freeze the O abundance at solar, since it would not have a significant influence on our spectral fitting above 2\,keV.  As a result, the model finds a best-fit iron abundance $A_{\rm Fe} = 5.5^{+2.5}_{-1.1}$, which is consistent with the measurement from \citealt{wes12}. This model provides a slightly better fit for the data ($\chi^{2}/$d.o.f.$=860.6/859$), with a higher coronal height $h = 4.70^{+4.04}_{-1.99}$, a less steep power-law continuum ($\Gamma = 1.79^{+0.17}_{-0.14}$) and a cutoff at $34^{+21}_{-10}$\,keV, even lower than in the previous model. As a result, the reflection fraction is reduced to $R_{\rm ref}=1.99$. We note here the cutoff energy is uncommonly low for an AGN, even compared with the lowest \nustar measurements \citep[e.g.,][]{urs15, tor16}. This model generally has more uncertainty in the parameters, most notably, places no constraint on the BH spin.  Although the metallicities in the whole galaxy should be related, there is no solid justification to assume the metallicities are the same in the BH vicinity and in the outer disk of the galaxy. However, we believe it is a reasonable simplification given the data quality, untying the iron abundances brings no obvious improvement to the fitting. 

Given the possible ionized absorption features in the Fe K band and motivated by the presence of large-scale outflows in the host galaxy \citep[e.g.,][]{wes12, bel13, rup15}, we add an ionized absorption component to account for any outflowing ionized absorbers. We modify the spectrum with an ionized absorption table model calculated by the {\small XSTAR} code \citep{kal01} and tie its iron abundance with that of the {\tt relxilllp} component. It is a physically reasonable requirement, since if the wind has extreme velocity, it most likely arises from the accretion disk, thus the same chemical abundances would naturally be expected.  We first assume the partial covering neutral absorber has a solar iron abundance (model~4a, see Table~\ref{tab:tab1}). The fitting results reveal an ionized absorber with a velocity $v_{\rm out} = 0.13^{+0.02}_{-0.06}\,c$ in the source rest-frame, which is well above the typical outflowing velocities of warm absorbers, about the common value found for UFOs \citep[e.g.,][]{tom10}. With an ionization parameter log$~\xi=3.06\pm0.07~\rm erg~cm~s^{-1}$, the absorption should be dominated by Fe XXV. However the detection of a high-velocity ionized outflow is not significant, as it only brings marginal improvement to the fitting $\Delta \chi^{2}/\Delta{\rm d.o.f} = -7.2/-3$ and the difference not visually evident in the ratio plot. Other best-fit parameters are consistent with model 3a (without the ionized absorber). A rapidly spinning BH with $a>0.89$ is required to provide an adequate fit for the data. 

We then reintroduce the neutral absorption model with a variable iron abundance (model 4b, see Table~\ref{tab:tab1}), and force the iron abundances of the disk reflection, the outflowing wind and the neutral absorber to be the same. The model brings no further improvement for the fitting. But again, it reduces the iron abundance $A_{\rm Fe}$ to $\sim$5, and since the iron abundance and the column density in the {\small XSTAR} model are degenerate parameters, the column density of the ionized outflow is estimated to be higher.  Best-fit parameters of the disk reflection agree well with those given by model 3b, except for a slightly higher inclination of $\theta = 51^{+4}_{-5}$~degrees. Only a low BH spin can be ruled out at the 90\% confidence level by this model.

\begin{figure}
\centering
\includegraphics[width=0.48\textwidth]{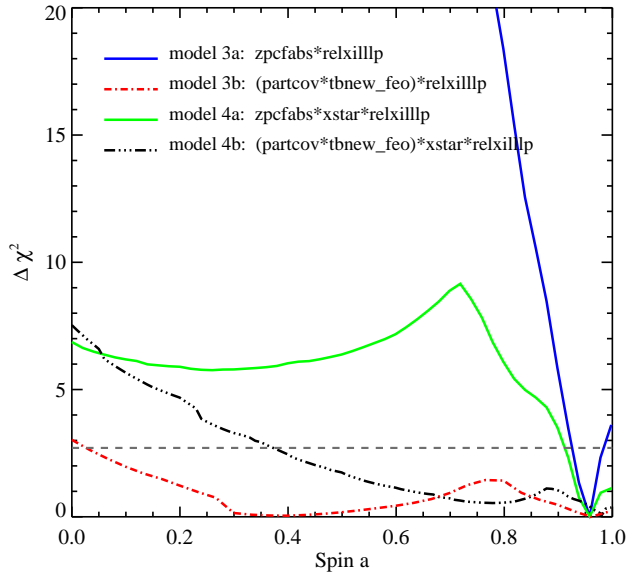}
\caption{$\Delta \chi^{2}$ as a function of BH spin for the relativistic disk reflection modeling under four different absorption scenarios. The dashed line marks the 90\% confidence limit. Model 3a and 4a favor a high BH spin of a $\gtrsim$ 0.9. The constraint weakens considerably when we consider a variable iron abundance for the neutral absorber in model 3b and 4b.
\label{fig:fig4}}
\end{figure}

\section{DISCUSSION}
\label{sec:dis}
We test four models under different absorption scenarios to physically fit for the disk reflection features in {IRAS\,05189--2524}, which provide good and comparable fits to the 2--30\,keV broadband spectrum (for details, see Table~\ref{tab:tab3} and  Figure~\ref{fig:fig3}). With the covering fraction of the neutral absorber fixed at 99\%, there is also no large discrepancy in the continuum below 2\,keV. Based on the limited S/N of the spectrum, we cannot distinguish which model better describes the data. A super-solar neutral absorber is a reasonable requirement by the metallicity measurement of the galaxy from the literature, and it reduces the disk iron abundance to a less extreme value. However, detailed modeling of the intrinsic absorption causes apparent difficulty in disentangling the parameter degeneracies. Adding an ionized absorber modeled by {\small XSTAR} helps to account for the possible features of a high-velocity outflow in the Fe K band, but is not statistically strongly required. For three out of the four models, we can rule out a slow or retrograde BH spin at the 90\% confidence level. In model 3a and 4a, a rapidly spinning BH is favored (for the spin constraints, see Figure~\ref{fig:fig4}). In addition, we consider the possible contribution from distant reflection produced in the torus or BLR, but including an extra {\tt xillver} component does not improve the fitting.

\subsection{Super-solar Iron Abundance}
In our spectral analysis of {IRAS\,05189--2524}, all acceptable fits require a super-solar iron abundance for the disk reflection component, but the value is not very well constrained. The primary effect of the iron abundance parameter is to change the relative strength of the iron line and the Compton hump. Because of the low net number counts at the high energy end of the \nustar spectra, the shape of the Compton hump is not well constrained, which could make it difficult to obtain a tight constraint of the iron abundance. We note that a similar over-abundance in iron has been reported in a number of AGNs \citep[e.g.,][]{ris09,pat11} and stellar-mass BHs \citep[e.g.,][]{gar15b}, with the examples of well-known narrow-line Seyfert 1 galaxies (NLSy1s) 1H\,0707--495 \citep{dau12,kar15} and IRAS\,13224--3809 \citep{fab13}. In the case of {IRAS\,05189--2524}, the galaxy is undergoing intense star formation, so it is expected to be iron enriched. Since super-solar metallicities were indicated via spatially resolved optical spectroscopy in the outer disk and the nuclear region of the galaxy \citep{wes12}, it is physically reasonable for the iron abundance to be high in the accretion disk.

\begin{figure}
\centering
\includegraphics[width=0.49\textwidth]{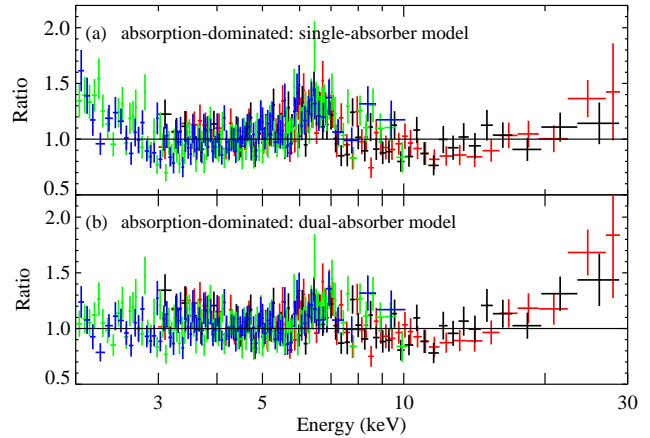}
\caption{Data/model ratio plots of the absorption-dominated models: (a) one neutral partial covering absorber with the covering fraction fixed at 99\% and a variable iron abundance; (b) two partial covering absorbers of free covering fractions and linked iron abundances. The data are rebinned for display clarity. 
\label{fig:fig5}}
\end{figure}

\subsection{Absorption-dominated Models}
Absorption from material lying relatively distant from the BH has frequently been proposed as an alternative interpretation of relativistic disk reflection \citep[e.g.,][]{mil08,mil09}. To investigate whether the excesses in the Fe K band and above 10\,keV observed in \iras can be fully accounted by absorption-dominated models (without the reflection component), we test modeling the 2--30\,keV spectrum with single and dual partial covering neutral absorption models with variable iron abundances. For the single-absorber model\footnote{(partcov*tbnew\_feo)*powerlaw}, we fix the covering faction at 99\% to avoid large deviation from the data below 2\,keV, the same way as in all the previous modeling. With $\Delta \chi^{2} = 215.6$ compared to model 3b, this scenario can be easily ruled out. The dual-absorber model\footnote{(partcov*tbnew\_feo)*(partcov*tbnew\_feo)*powerlaw} with an iron abundance $A_{\rm Fe}=2.2^{+0.2}_{-0.4}$ provides a better fit for the data, which requires one full covering absorber with the column density $N_{\rm H,abs\,1}=6.01^{+0.14}_{-0.93}\times10^{22}$~cm$^{-2}$ and another of the coverage fraction $f_{\rm abs\,2}=64\pm5$\% with $N_{\rm H,abs\,2}=1.92^{+0.58}_{-0.18}\times10^{23}$~cm$^{-2}$. However, it is still statistically worse than the reflection-dominated model with $\Delta \chi^{2}=75.6$. The model leaves a broad excess in the Fe K band and also fails to produce the shape of the Compton hump above 10\,keV (see Figure~\ref{fig:fig4}, panel b), resembling the case in NGC 1365 \citep{ris13} .

\subsection{Eddington Ratio and Outflow} 
The time-averaged 2--30\,keV X-ray flux observed by \nustar is $7.1 \times 10^{-12}$~ergs~cm$^{-2}$~s$^{-1}$, corresponding to an intrinsic luminosity of $3.0 \times 10^{43}$~ergs~s$^{-1}$ during the 2013 observation. Using the central velocity dispersions measured from the Ca {\footnotesize II} Triplet line widths \citep{rot13} and the $M_{\rm BH}$-$\sigma$ relation \citep{tre02}, the BH mass of \iras is estimated to be $M_{\rm BH}=4.2 \times10^{8}~\Msun$. We note the BH mass derived from the Ca {\footnotesize II} Triplet here is more than ten times larger than the value measured from CO band heads in the near infrared \citep{das06}. As discussed in \cite{das06}, CO band heads are often representative of young stellar populations, which could lead to systematically lower BH masses for ULIRGs. Combined with the bolometric luminosity inferred from the infrared \citep{ten15}, \iras is measured to be accreting at an Eddington rate $\lambda_{\rm Edd}=L_{\rm bol}/L_{\rm Edd}$ of 0.12. Based on the correlation between the photon indexes and the Eddington ratios of AGNs \citep{she08,bri13,bri16}, a photon index of $\Gamma \sim 2.0$ would be expected. The photon index obtained by our disk reflection modeling with a low cutoff energy is $\Gamma \simeq1.7-2.0$, which is consistent with the expected value given the scatter of the relation.

In our spectral modeling, we did not detect significant disk winds. Fitting the possible absorption features in the spectrum with a simple Gaussian absorption line using the {\tt gabs} model would reach similar $\chi^{2}$ as the physical {\small XSTAR} grid, with the best-fit line centroid at $\sim$7.33\,keV.  Although the effect of this ionized absorber on the continuum is subtle, we include it in the spectral fitting, as even mild absorption lying around the Fe K-edge could influence the BH spin measurement, which is sensitive to the profile of the broad iron line. The power of an outflow scales with the amount of matter being ejected \citep{kin15}. With the rather small column density measured for the ionized absorption compared with that of massive disk winds found in other AGNs \citep[e.g., IRAS\,11119+3257,][]{tom15}, even if an UFO indeed exist in {IRAS\,05189--2524} at the epoch of the observation, it is probably not energetic enough to to have a large impact on its galactic environment. 

\section{SUMMARY AND CONCLUSION}
\label{sec:con}
We detect disk reflection features of a broad iron line and a Compton reflection hump in {IRAS\,05189--2524}, a Seyfert 1h galaxy and ULIRG, from the \nustar and \xmm data taken in 2013. With the broadband X-ray spectrum, an alternative dual partial covering absorber explanation can be ruled out. Physically modeling with the self-consistent relativistic reflection model {\tt relxilllp} finds that the reflection is likely to be generated from the inner range of a rapidly spinning BH. The high reflection fraction measured ($R_{\rm ref} \simeq$ 2.0--3.2) indicates the reflection area is close to the BH, where relativistic light bending effects boost the reflection signature \citep{min04}, suggesting the coronal illuminating source is compact. However, due to the complex effects from neutral and ionized absorption and the limited S/N data, parameter degeneracies preclude putting a tight, model-independent estimation on the BH spin. 

As a merger system, \iras is an interesting target for the spin measurement, as BH spins are believed to encode information about the galaxy evolution history \citep{ber08}, and ULIRGs have been proposed as an important transition phase in the hierarchical evolution model of galaxies \citep{hop06, san88}. Simulations predict it is uncommon for mergers to yield large spins, unless alignment of the spins of the merging BHs with the orbital angular momentum is very efficient \citep{ber08}. 

With Fe K band absorption features marginally evident in the residuals, the detection of a high-velocity ionized wind is not statistically significant. Also, we find the iron abundance of the neural absorber is strongly degenerated with the BH spin. Future observations with higher S/N data could help disentangle the reflection component from the underlying continuum and the various absorption effects. Assuming the source remains at the flux level observed in 2013, if the ionized absorption indeed exists in IRAS\,05189--2524, simulations indicate the XSTAR component would be significantly required by datasets with more than $\sim$2 times of exposure, leading to a $>3 \sigma$ detection of the ionized absorption features. With longer observations (500 ks \nustar and 300 ks \xmm exposures), it would be possible to obtain the BH spin to within $\sim$30\% of uncertainty under all the different absorption scenarios we have discussed above. 

\acknowledgments{
We thank the anonymous referee for helpful comments that improved the paper. M.\,B. acknowledges support from NASA Headquarters under the NASA Earth and Space Science Fellowship Program, grant NNX14AQ07H. This work was supported under NASA Contract No.~NNG08FD60C, and made use of data obtained with NuSTAR, a project led by Caltech, funded by NASA and managed by NASA/JPL and has utilized the NUSTARDAS software package, jointly developed by the ASDC (Italy) and Caltech (USA). We thank the \nustar Operations, Software and  Calibration teams for support with the execution and analysis of these observations. This research has also made use of data obtained with XMM-Newton, an ESA science mission with instruments and contributions directly funded by ESA Member States.
}

\bibliographystyle{yahapj}

\begin{thebibliography}{}

\bibitem[Arnaud(1996)]{arn96} 
Arnaud, K.~A.\ 1996, Astronomical Data Analysis Software and Systems V, 101, 17 

\bibitem[Bellocchi et al.(2013)]{bel13}
 Bellocchi, E., Arribas, S., Colina, L., \& Miralles-Caballero, D.\ 2013, \aap, 557, A59 
 
 \bibitem[Berti \& Volonteri(2008)]{ber08} 
 Berti, E., \& Volonteri, M.\ 2008, \apj, 684, 822-828 

\bibitem[Brightman et al.(2016)]{bri16}
Brightman, M., Masini, A., Ballantyne, D.~R., et al.\ 2016, \apj, 826, 93

\bibitem[Brightman \& Nandra(2011)]{bri11}
Brightman, M., \& Nandra, K.\ 2011, \mnras, 413, 1206 

\bibitem[Brightman et al.(2013)]{bri13} 
Brightman, M., Silverman, J.~D., Mainieri, V., et al.\ 2013, \mnras, 433, 2485 

\bibitem[Dasyra et al.(2006)]{das06} 
Dasyra, K.~M., Tacconi, L.~J., Davies, R.~I., et al.\ 2006, \apj, 651, 835 

\bibitem[Dauser et al.(2014)]{dau14}
Dauser, T., Garc{\'{\i}}a, J., Parker, M.~L., Fabian, A.~C., \& Wilms, J.\ 2014, \mnras, 444, L100 

\bibitem[Dauser et al.(2013)]{dau13} 
Dauser, T., Garcia, J., Wilms, J., et al.\ 2013, \mnras, 430, 1694 

\bibitem[Dauser et al.(2012)]{dau12}
Dauser, T., Svoboda, J., Schartel, N., et al.\ 2012, \mnras, 422, 1914 

\bibitem[Dewangan et al.(2003)]{dew03}
Dewangan, G.~C., Griffiths, R.~E., \& Schurch, N.~J.\ 2003, \apj, 592, 52 

\bibitem[Fabian et al.(2013)]{fab13} 
Fabian, A.~C., Kara, E., Walton, D.~J., et al.\ 2013, \mnras, 429, 2917

\bibitem[Fabian et al.(1989)]{fab89} 
Fabian, A.~C., Rees, M.~J., Stella, L., \& White, N.~E.\ 1989, \mnras, 238, 729  

\bibitem[Feruglio et al.(2015)]{fer15} 
Feruglio, C., Fiore, F., Carniani, S., et al.\ 2015, \aap, 583, A99 

\bibitem[Garc{\'{\i}}a et al.(2014)]{gar14} 
Garc{\'{\i}}a, J., Dauser, T., Lohfink, A., et al.\ 2014, \apj, 782, 76 

\bibitem[Garc{\'{\i}}a et al.(2015a)]{gar15a}
Garc{\'{\i}}a, J.~A., Dauser, T., Steiner, J.~F., et al.\ 2015, \apjl, 808, L37 

\bibitem[Garc{\'{\i}}a et al.(2015b)]{gar15b} 
Garc{\'{\i}}a, J.~A., Steiner, J.~F., McClintock, J.~E., et al.\ 2015, \apj, 813, 84 

\bibitem[George \& Fabian(1991)]{geo91} 
George, I.~M., \& Fabian, A.~C.\ 1991, \mnras, 249, 352

\bibitem[Guainazzi et al.(2010)]{gua10} 
Guainazzi, M., Bianchi, S., Matt, G., et al.\ 2010, \mnras, 406, 2013 

\bibitem[Haardt \& Maraschi(1993)]{haa93}
Haardt, F., \& Maraschi, L.\ 1993, \apj, 413, 507

\bibitem[Harrison et al.(2013)]{har13} 
Harrison, F.~A., Craig, W.~W., Christensen, F.~E., et al.\ 2013, \apj, 770, 103

\bibitem[Hopkins et al.(2006)]{hop06} 
Hopkins, P.~F., Hernquist, L., Cox, T.~J., et al.\ 2006, \apjs, 163, 1 

\bibitem[Jansen et al.(2001)]{jan01} 
Jansen, F., Lumb, D., Altieri, B., et al.\ 2001, \aap, 365, L1

\bibitem[Kallman \& Bautista(2001)]{kal01} 
Kallman, T., \& Bautista, M.\ 2001, \apjs, 133, 221 

\bibitem[Kara et al.(2015)]{kar15}
Kara, E., Fabian, A.~C., Lohfink, A.~M., et al.\ 2015, \mnras, 449, 234 

 \bibitem[Iwasawa et al.(1996)]{iwa96} 
 Iwasawa, K., Fabian, A.~C., Mushotzky, R.~F., et al.\ 1996, \mnras, 279, 837 
 
\bibitem[Kalberla et al.(2005)]{kal05} 
Kalberla, P.~M.~W., Burton, W.~B., Hartmann, D., et al.\ 2005, \aap, 440, 775  

\bibitem[King \& Pounds(2015)]{kin15} 
King, A., \& Pounds, K.\ 2015, \araa, 53, 115 
 
\bibitem[Laor(1991)]{lao91} 
Laor, A.\ 1991, \apj, 376, 90 

\bibitem[Lehmer et al.(2010)]{leh10} 
Lehmer, B.~D., Alexander, D.~M., Bauer, F.~E., et al.\ 2010, \apj, 724, 559 

\bibitem[Madsen et al.(2015)]{mad15}
 Madsen, K.~K., Harrison, F.~A., Markwardt, C.~B., et al.\ 2015, \apjs, 220, 8 

\bibitem[Magdziarz \& Zdziarski(1995)]{mag95} 
Magdziarz, P., \& Zdziarski, A.~A.\ 1995, \mnras, 273, 837 

\bibitem[Markoff et al.(2005)]{mar05} 
Markoff, S., Nowak, M.~A., \& Wilms, J.\ 2005, \apj, 635, 1203 

\bibitem[Matt et al.(1992)]{mat92} 
Matt, G., Perola, G.~C., Piro, L., \& Stella, L.\ 1992, \aap, 257, 63 

\bibitem[Miller et al.(2008)]{mil08} 
Miller, L., Turner, T.~J., \& Reeves, J.~N.\ 2008, \aap, 483, 437 

\bibitem[Miller et al.(2009)]{mil09} 
Miller, L., Turner, T.~J., \& Reeves, J.~N.\ 2009, \mnras, 399, L69 

\bibitem[Miniutti \& Fabian(2004)]{min04} 
Miniutti, G., \& Fabian, A.~C.\ 2004, \mnras, 349, 1435 

\bibitem[Miniutti et al.(2007)]{min07}
Miniutti, G., Ponti, G., Dadina, M., Cappi, M., \& Malaguti, G.\ 2007, \mnras, 375, 227 

\bibitem[Murphy \& Yaqoob(2009)]{mur09}
Murphy, K.~D., \& Yaqoob, T.\ 2009, \mnras, 397, 1549 

\bibitem[Nandra et al.(2007)]{nan07} 
Nandra, K., O'Neill, P.~M., George, I.~M., \& Reeves, J.~N.\ 2007, \mnras, 382, 194

\bibitem[Nandra \& Pounds(1996)]{nan94}
Nandra, K., \& Pounds, K.~A.\ 1994, \mnras, 268, 405

\bibitem[Nardini et al.(2015)]{nar15} 
Nardini, E., Reeves, J.~N., Gofford, J., et al.\ 2015, Science, 347, 860 

\bibitem[Parker et al.(2014)]{par14}
Parker, M.~L., Wilkins, D.~R., Fabian, A.~C., et al.\ 2014, \mnras, 443, 1723 

\bibitem[Patrick et al.(2011)]{pat11} 
Patrick, A.~R., Reeves, J.~N., Lobban, A.~P., Porquet, D., \& Markowitz, A.~G.\ 2011, \mnras, 416, 2725 

\bibitem[Pounds et al.(2003)]{pou03} 
Pounds, K.~A., King, A.~R., Page, K.~L., \& O'Brien, P.~T.\ 2003, \mnras, 346, 1025  

\bibitem[Pounds \& King(2013)]{pou13} 
Pounds, K.~A., \& King, A.~R.\ 2013, \mnras, 433, 1369 
  
\bibitem[Ricci et al.(2014)]{ric14} 
Ricci, C., Tazaki, F., Ueda, Y., et al.\ 2014, \apj, 795, 147 

\bibitem[Risaliti et al.(2013)]{ris13} 
Risaliti, G., Harrison, F.~A., Madsen, K.~K., et al.\ 2013, \nat, 494, 449 

\bibitem[Risaliti et al.(2009)]{ris09}
 Risaliti, G., Miniutti, G., Elvis, M., et al.\ 2009, \apj, 696, 160 

\bibitem[Rivers et al.(2013)]{riv13}
Rivers, E., Markowitz, A., \& Rothschild, R.\ 2013, \apj, 772, 114

\bibitem[Rothberg et al.(2013)]{rot13} 
Rothberg, B., Fischer, J., Rodrigues, M., \& Sanders, D.~B.\ 2013, \apj, 767, 72 

\bibitem[Rupke \& Veilleux(2015)]{rup15} 
Rupke, D.~S.~N., \& Veilleux, S.\ 2015, \apj, 801, 126 

\bibitem[Rupke et al.(2008)]{rup08} 
Rupke, D.~S.~N., Veilleux, S., \& Baker, A.~J.\ 2008, \apj, 674, 172-193 

\bibitem[Rupke et al.(2005)]{rup05} 
Rupke, D.~S., Veilleux, S., \& Sanders, D.~B.\ 2005, \apj, 632, 751

\bibitem[Sanders et al.(1988)]{san88} 
Sanders, D.~B., Soifer, B.~T., Elias, J.~H., et al.\ 1988, \apj, 325, 74 

\bibitem[Severgnini et al.(2001)]{sev01} 
Severgnini, P., Risaliti, G., Marconi, A., Maiolino, R., \& Salvati, M.\ 2001, \aap, 368, 44 
 
\bibitem[Shemmer et al.(2008)]{she08}
Shemmer, O., Brandt, W.~N., Netzer, H., Maiolino, R., \& Kaspi, S.\ 2008, \apj, 682, 81-93 
  
\bibitem[Str{\"u}der et al.(2001)]{str01} 
Str{\"u}der, L., Briel, U., Dennerl, K., et al.\ 2001, \aap, 365, L18 

\bibitem[Tan et al.(2012)]{tan12} 
 Tan, Y., Wang, J.~X., Shu, X.~W., \& Zhou, Y.\ 2012, \apjl, 747, L11 
 
\bibitem[Tanaka et al.(1995)]{tan95} 
Tanaka, Y., Nandra, K., Fabian, A.~C., et al.\ 1995, \nat, 375, 659 

\bibitem[Teng et al.(2015)]{ten15} 
Teng, S.~H., Rigby, J.~R., Stern, D., et al.\ 2015, \apj, 814, 56 

\bibitem[Teng et al.(2009)]{ten09} 
Teng, S.~H., Veilleux, S., Anabuki, N., et al.\ 2009, \apj, 691, 261 

\bibitem[Teng et al.(2013)]{ten13} 
Teng, S.~H., Veilleux, S., \& Baker, A.~J.\ 2013, \apj, 765, 95 

\bibitem[Tombesi et al.(2010)]{tom10}
 Tombesi, F., Cappi, M., Reeves, J.~N., et al.\ 2010, \aap, 521, A57 
 
\bibitem[Tombesi et al.(2015)]{tom15}
  Tombesi, F., Mel{\'e}ndez, M., Veilleux, S., et al.\ 2015, \nat, 519, 436 
  
\bibitem[Tortosa et al.(2016)]{tor16} 
Tortosa, A., Marinucci, A., Matt, G., et al.\ 2016, arXiv:1612.05871 
  
\bibitem[Tremaine et al.(2002)]{tre02} 
Tremaine, S., Gebhardt, K., Bender, R., et al.\ 2002, \apj, 574, 740 

\bibitem[Turner et al.(2001)]{tur01} 
 Turner, M.~J.~L., Abbey, A., Arnaud, M., et al.\ 2001, \aap, 365, L27
 
 \bibitem[U et al.(2012)]{u12} 
 U, V., Sanders, D.~B., Mazzarella, J.~M., et al.\ 2012, \apjs, 203, 9 
 
 \bibitem[Ursini et al.(2015)]{urs15}
  Ursini, F., Boissay, R., Petrucci, P.-O., et al.\ 2015, \aap, 577, A38 

\bibitem[Veilleux et al.(1995)]{vei95}
 Veilleux, S., Kim, D.-C., Sanders, D.~B., Mazzarella, J.~M., \& Soifer, B.~T.\ 1995, \apjs, 98, 171 

\bibitem[Veilleux et al.(1999a)]{vei99a} 
Veilleux, S., Kim, D.-C., \& Sanders, D.~B.\ 1999a, \apj, 522, 113

\bibitem[Veilleux et al.(2013)]{vei13} 
Veilleux, S., Mel{\'e}ndez, M., Sturm, E., et al.\ 2013, \apj, 776, 27

\bibitem[Veilleux et al.(1999b)]{vei99b} 
Veilleux, S., Sanders, D.~B., \& Kim, D.-C.\ 1999b, \apj, 522, 139  

\bibitem[V{\'e}ron-Cetty \& V{\'e}ron(2006)]{ver06} 
V{\'e}ron-Cetty, M.-P., \& V{\'e}ron, P.\ 2006, \aap, 455, 773 

\bibitem[Walton et al.(2013)]{wal13} 
Walton, D.~J., Nardini, E., Fabian, A.~C., Gallo, L.~C., \& Reis, R.~C.\ 2013, \mnras, 428, 2901 

\bibitem[Walton et al.(2014)]{wal14}
Walton, D.~J., Risaliti, G., Harrison, F.~A., et al.\ 2014, \apj, 788, 76

\bibitem[Westmoquette et al.(2012)]{wes12} 
Westmoquette, M.~S., Clements, D.~L., Bendo, G.~J., \& Khan, S.~A.\ 2012, \mnras, 424, 416 

\bibitem[Wilms et al.(2000)]{wil00} 
 Wilms, J., Allen, A., \& McCray, R.\ 2000, \apj, 542, 914


\bibitem[Winter et al.(2009)]{win09}
Winter, L.~M., Mushotzky, R.~F., Reynolds, C.~S., \& Tueller, J.\ 2009, \apj, 680, 1322

\bibitem[Young et al.(1996)]{you96} 
Young, S., Hough, J.~H., Efstathiou, A., et al.\ 1996, \mnras, 281, 1206 

\end{thebibliography}

\end{document}